\documentclass{article}
\usepackage{spconf,amsmath,graphicx,hyperref}
\usepackage{algorithm}
\usepackage{algorithmicx}
\usepackage{algpseudocode} % 伪代码格式更漂亮
\usepackage{amsmath}       % 数学公式
\usepackage{amssymb}
\usepackage{booktabs}
\usepackage[table]{xcolor}

\title{Agent-GSPO: Communication-Efficient Multi-Agent Systems via Group Sequence Policy Optimization}
\name{Yijia Fan; Jusheng Zhang; Jing Yang; Keze Wang}
\address{Sun Yat-sen University}
\begin{document}
%\ninept
%
\maketitle
\begin{abstract}
To combat the prohibitive communication costs of ``free-for-all" multi-agent systems (MAS), we introduce \textbf{Agent-GSPO}, a framework that directly optimizes for token economy using sequence-level reinforcement learning. Agent-GSPO leverages the stable and memory-efficient Group Sequence Policy Optimization (GSPO) algorithm to train agents on a communication-aware reward that explicitly penalizes verbosity. Across seven reasoning benchmarks, Agent-GSPO not only achieves new state-of-the-art performance but does so with a fraction of the token consumption of existing methods. By fostering emergent strategies like ``strategic silence," our approach provides a practical blueprint for developing scalable and economically viable multi-agent systems.
\end{abstract}
\section{Introduction}
\label{sec:intro}

The research frontier for Large Language Models (LLMs)\cite{gpt2,Z2} is rapidly shifting from single-agent solvers to complex Multi-Agent Systems (MAS)\cite{mas1,Z2,Z7}. This paradigm shift holds the promise of tackling problems far beyond the reach of any individual model. This collaborative approach has demonstrated significant potential in domains like complex reasoning and software development, where agents can pool distributed knowledge and refine solutions through iterative dialogue.

However, this collaborative promise is frequently undermined by a critical bottleneck: communication inefficiency. Many advancements\cite{mas2,Z4} rely on ``free-for-all" communication protocols, permitting agents to broadcast information with little to no cost. This architectural choice often leads to an exponential increase in token consumption and a low signal-to-noise ratio, inundating agents with verbose, low-value exchanges. Existing methods\cite{du2023improvingfactualityreasoninglanguage,ext2,Z3}, despite their performance gains, often overlook this severe resource cost, resulting in prohibitive operational overhead that hinders practical, large-scale deployment. We argue that the core issue is not a lack of communication, but the absence of a principled mechanism to enforce communicative discipline and resource rationality.

To address this challenge, we turn to sequence-level reinforcement learning (RL), a paradigm naturally suited for optimizing the generation of entire messages. Specifically, we leverage Group Sequence Policy Optimization (GSPO)\cite{gspo}, a recent advancement that stabilizes policy updates by operating on groups of entire response sequences rather than individual tokens. Unlike token-level objectives common in Proximal Policy Optimization (PPO)\cite{ppo} applications, GSPO's sequence-level clipping mechanism is inherently more stable for optimizing long, variable-length action sequences. This makes it an ideal framework for multi-agent dialogue, where an agent's ``action" is a complete utterance. Furthermore, by directly optimizing the policy on sequence-level rewards, GSPO avoids the need for an auxiliary reward or value model for every token, leading to greater memory efficiency during training.

Building on this foundation, we introduce \textbf{Agent-GSPO}, a novel framework designed to cultivate communication efficiency in LLM-based MAS. Agent-GSPO frames the generation of a message as a sequential action and directly optimizes a communication-aware reward function. This reward signal explicitly incentivizes task success while penalizing communication overhead, such as token count and conversational turns. By applying the GSPO objective, our framework encourages agents to learn a sophisticated trade-off: to ``speak less but more precisely." This process intrinsically prunes low-value chatter and fosters the emergent skill of ``strategic silence," where agents learn to withhold information that is redundant or unlikely to contribute to the team's success.

Extensive experiments demonstrate that Agent-GSPO achieves new state-of-the-art performance across several challenging reasoning benchmarks. Crucially, this is accomplished with remarkable efficiency, consuming only a fraction of the communication tokens used by current state-of-the-art methods. Our analysis reveals that Agent-GSPO effectively learns to adapt its communication strategies, shifting from verbosity to conciseness based on the implicit value of its messages, thereby establishing a more practical and scalable framework for multi-agent collaboration.

\begin{figure*}[t]
    \centering
    \includegraphics[width=0.9\linewidth]{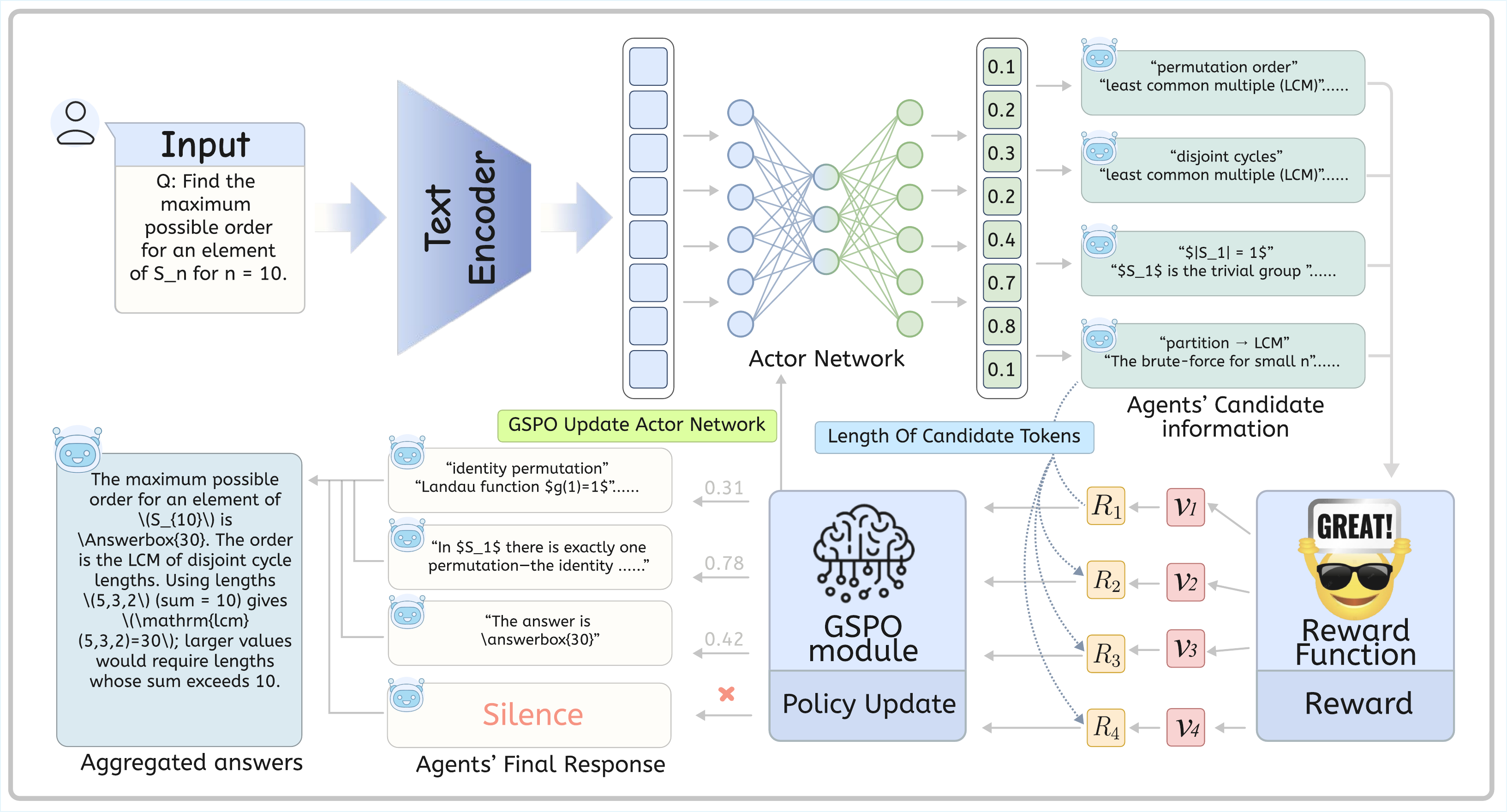}
    \caption{
    Overview of the Agent-GSPO training pipeline. 
    The Actor Network (\(\pi_{\theta_{\text{old}}}\)) samples a group of candidate responses. 
    Each response receives a communication-aware reward (\(r\)), which balances task success (\(r_{\text{task}}\)) against a token-length penalty. 
    The GSPO module processes these sequence-level rewards to update the Actor's policy (\(\pi_\theta\)), training the agent to favor concise and effective communication.
    }
    \label{fig:placeholder}
\end{figure*}

\section{Related Work}
\label{sec:related_work}

\textbf{Communication in Multi-Agent Systems.} The challenge of efficient communication in Multi-Agent Systems (MAS) has evolved significantly with the advent of LLMs. Early research focused on structured protocols like Agent Communication Languages (ACLs) and negotiation frameworks like the Contract Net Protocol to ensure clarity and manage overhead. More recently, the focus has shifted to learning emergent communication strategies. However, the ``free-for-all" nature of LLM-based dialogue often leads to prohibitive token costs. To address this, current methods have explored various pruning strategies, such as gating mechanisms that decide when an agent should speak, or topological pruning that removes communication links, as seen in AgentPrune-R \cite{AgentPrune-R,Z8}. Our work diverges from these approaches by not relying on external pruning or auction mechanisms. Instead, we formulate the problem as a direct policy optimization task, where agents intrinsically learn to be concise by optimizing a communication-aware reward signal.

\textbf{Reinforcement Learning for LLMs.} Reinforcement Learning (RL) has become a cornerstone for aligning LLMs with desired behaviors\cite{rlhf}, with Proximal Policy Optimization (PPO) being the de-facto standard for both post-training alignment and tool use. However, PPO's token-level objective can suffer from high variance and instability when optimizing long, sequential actions like natural language messages. Furthermore, its reliance on a separate critic or value network adds significant memory and computational overhead. To overcome these limitations, recent work has introduced sequence-level policy optimization methods. Group Sequence Policy Optimization (GSPO), the algorithm we employ, stabilizes training by computing a clipped importance sampling ratio over entire sequences and uses group-wise reward normalization to estimate advantages directly, obviating the need for a critic network. 

\section{Agent-GSPO}

\subsection{Communication-Aware Reward Function}
Given an input query $x$, the policy $\pi_\theta$ generates a response sequence $y = (y_1, \dots, y_{|y|})$. We define a composite reward function that balances task accuracy with communication efficiency:
\begin{equation}
\begin{split}
r(x,y) = r_{\text{task}}(x,y) 
&- \lambda_{\text{tok}} \cdot \text{tokens}(y) \\
&- \lambda_{\text{turn}} \cdot \text{turns}(y)
- \lambda_{\text{rep}} \cdot \text{repetition}(y),
\end{split}
\label{eq:reward}
\end{equation}
where $r_{\text{task}}$ measures task-specific performance (e.g., exact match, pass@1). The remaining terms penalize token usage, conversational turns, and content repetition. The coefficient $\lambda_{\text{tok}}$ is the primary hyperparameter controlling token efficiency.

\subsection{Sequence-Level GSPO Objective}
To ensure optimization stability, GSPO operates at the sequence level. For each input $x$, we sample a group of $G$ candidate responses $\{y_i\}_{i=1}^G$ from a frozen, older policy $\pi_{\theta_{\text{old}}}$. We then compute group-wise normalized advantages:
\begin{equation}
\hat{A}_i = \frac{r(x,y_i) - \mu_r}{\sigma_r + \epsilon},
\end{equation}
where $\mu_r = \frac{1}{G}\sum_{j=1}^G r(x,y_j)$ is the mean reward and $\sigma_r$ is the standard deviation of rewards within the group, with $\epsilon$ as a small constant for numerical stability. To mitigate variance arising from sequence length disparities, the importance sampling ratio is normalized by the sequence length:
\begin{equation}
\begin{split}
s_i(\theta) &= 
\Bigg(\frac{\pi_\theta(y_i|x)}{\pi_{\theta_{\text{old}}}(y_i|x)}\Bigg)^{\frac{1}{|y_i|}} \\
&= \exp\!\Bigg( \frac{1}{|y_i|}\sum_{t=1}^{|y_i|} \log \frac{\pi_\theta(y_{i,t}|x,y_{i,<t})}{\pi_{\theta_{\text{old}}}(y_{i,t}|x,y_{i,<t})} \Bigg).
\end{split}
\label{eq:seq-ratio}
\end{equation}
The final GSPO objective, analogous to PPO, incorporates a clipping mechanism at the sequence level:

\begin{equation}
\begin{aligned}
J_{\text{GSPO}}(\theta) = \mathbb{E}_{x, \{y_i\} \sim \pi_{\theta_{\text{old}}}} &\Bigg[ \frac{1}{G}\sum_{i=1}^G \min\Big(  s_i(\theta)\hat{A}_i, \\
& \text{clip}(s_i(\theta), 1-\varepsilon, 1+\varepsilon)\hat{A}_i \Big) \Bigg].
\end{aligned}
\end{equation}
\label{eq:gspo-obj}

This sequence-level clipping is crucial for stabilizing training; it prevents the large-variance updates that can arise from token-level probability ratios, especially for long sequences.

\subsection{Dual Budget Constraint}
To explicitly enforce an expected communication budget, $B$, we can treat $\lambda_{\text{tok}}$ as a learnable dual variable, updated via projected gradient ascent:
\begin{equation}
\lambda_{\text{tok}} \leftarrow 
\max \big( 0, \lambda_{\text{tok}} + \eta \cdot (\overline{\text{tokens}} - B) \big),
\end{equation}
where $\overline{\text{tokens}}$ is the batch-average token consumption and $\eta$ is a learning rate. This mechanism dynamically adjusts the penalty to steer the average communication cost towards the target budget.

\subsection{Training Algorithm}
The training procedure is summarized in Algorithm~\ref{alg:Agent-GSPO}.
\begin{algorithm}[h]
\caption{Agent-GSPO Training}
\label{alg:Agent-GSPO}
\begin{algorithmic}[1]
\State Initialize policy $\pi_\theta$, old policy $\pi_{\theta_{\text{old}}}$, and penalty $\lambda_{\text{tok}}$.
\For{each training iteration}
    \State Sample a batch of queries $\{x_b\}_{b=1}^B$.
    \For{each query $x_b$}
        \State Sample $G$ responses $\{y_i\}_{i=1}^G \sim \pi_{\theta_{\text{old}}}( \cdot | x_b)$.
        \State Compute rewards $r(x_b,y_i)$ using Eq.~\ref{eq:reward}.
        \State Compute group-wise normalized advantages $\hat{A}_i$.
    \EndFor
    \State Compute sequence-level importance ratios $s_i(\theta)$ via Eq.~\ref{eq:seq-ratio}.
    \State Construct the GSPO objective $J_{\text{GSPO}}$ as in Eq.~\ref{eq:gspo-obj}.
    \State Update policy parameters: $\theta \leftarrow \theta + \alpha \nabla_\theta J_{\text{GSPO}}$.
    \State Periodically update the old policy: $\pi_{\theta_{\text{old}}} \leftarrow \pi_\theta$.
    \State \textbf{if} using dual budget constraint \textbf{then} update $\lambda_{\text{tok}}$.
\EndFor
\end{algorithmic}
\end{algorithm}

\subsection{Emergent Communication Efficiency}
Through this optimization, agents learn to intrinsically balance task performance and communication cost. As $\lambda_{\text{tok}}$ increases (either manually or via the dual update), the policy shifts from verbose outputs towards concise summaries, keywords, or even abstains from communicating altogether. This emergent ``strategic silence'' is a key benefit of our framework, enabling state-of-the-art performance with substantially lower communication overhead.

\begin{table*}[t]
  \centering
  \caption{Overall performance comparison across seven reasoning benchmarks. 
  We report accuracy (\%) for all tasks except HumanEval, for which we report pass@1 (\%). 
  The best score in each column is in \textbf{bold}. Values in parentheses show absolute gain over the Vanilla baseline.}
  \label{tab:main_performance_updated}
  %––––– 基本排版微调 –––––%
  \footnotesize
  \setlength\tabcolsep{3pt}

  %––––– 自动缩放至整页宽 –––––%
  \resizebox{\textwidth}{!}{%
  \begin{tabular}{|l||c|c|c|c|c|c|c|}
    \hline
    \textbf{Method} & \textbf{MMLU} & \textbf{MultiArith} & \textbf{GSM8K} & \textbf{SVAMP} &
    \textbf{AQuA} & \textbf{HumanEval} & \textbf{MATH-500} \\
    \hline\hline
    \multicolumn{8}{|l|}{\textit{Single-Agent Methods}} \\ \hline
    Vanilla        & 82.14 & 93.15 & 85.40 & 87.18 & 70.34 & 71.68 & 73.72 \\
    CoT            & 82.65 (+0.51) & 94.79 (+1.64) & 87.17 (+1.77) & 88.32 (+1.14) &
                     73.91 (+3.57) & 75.52 (+3.84) & 75.18 (+1.46) \\
    ComplexCoT     & 83.78 (+1.64) & 95.86 (+2.71) & 87.62 (+2.22) & 90.17 (+2.99) &
                     77.58 (+7.24) & 74.94 (+3.26) & 76.85 (+3.13) \\
    SC             & 82.66 (+0.52) & 96.88 (+3.73) & 87.93 (+2.53) & 88.69 (+1.51) &
                     75.08 (+4.74) & 77.30 (+5.62) & 77.02 (+3.30) \\
    \hline
    \multicolumn{8}{|l|}{\textit{Multi-Agent Methods}} \\ \hline
    PHP            & 83.45 (+1.31) & 96.41 (+3.26) & 92.45 (+7.05) & 90.62 (+3.44) &
                     76.25 (+5.91) & 82.96 (+11.28) & 79.24 (+5.52) \\
    LLM-Debate     & 83.69 (+1.55) & 96.27 (+3.12) & 90.23 (+4.83) & 90.56 (+3.38) &
                     77.52 (+7.18) & 83.79 (+12.11) & 80.15 (+6.43) \\
    DyLAN          & 80.16 (-1.98) & 94.27 (+1.12) & 88.16 (+2.76) & 87.40 (+0.22) &
                     74.16 (+3.82) & 89.70 (+18.02) & 81.66 (+7.94) \\
    AgentPrune-R   & 83.94 (+1.80) & 96.30 (+3.15) & 95.83 (+10.43) & 91.68 (+4.50) &
                     78.60 (+8.26) & 90.30 (+18.62) & 82.81 (+9.09) \\
    \hline
    \rowcolor{gray!22}
    \textbf{Agent-GSPO} &
    \textbf{84.10 (+1.96)} & \textbf{97.20 (+4.05)} & \textbf{96.02 (+10.62)} &
    \textbf{91.90 (+4.72)} & \textbf{79.80 (+9.46)} & \textbf{90.70 (+19.02)} &
    \textbf{83.10 (+9.38)} \\
    \hline
  \end{tabular}%
 } % end resizebox
\end{table*}

\begin{table}[h!]
\centering
\caption{Consolidated performance comparison on MMLU and GSM8K datasets. Our Agent-GSPO, highlighted in gray, demonstrates superior or competitive performance with significantly better token efficiency.}
\label{tab:combined_results}
% 使用 resizebox 可以确保表格宽度不超过页面文本宽度
\setlength\tabcolsep{1pt}
\resizebox{\linewidth}{!}{%
\begin{tabular}{l cc cc}
\toprule
& \multicolumn{2}{c}{\textbf{MMLU}} & \multicolumn{2}{c}{\textbf{GSM8K}} \\
\cmidrule(lr){2-3} \cmidrule(lr){4-5}
\textbf{Method} & \textbf{Accuracy (\%)} & \textbf{Token Cons.} & \textbf{Accuracy (\%)} & \textbf{Token Cons.} \\
\midrule
\rowcolor{lightgray} \textbf{Agent-GSPO}        & 84.1 & $9.20 \times 10^5$ & 96.0 & $7.20 \times 10^6$ \\
LLM-Debate        & 83.7 & $1.50 \times 10^6$ & 90.2 & $2.20 \times 10^7$ \\
PHP               & 83.4 & $2.60 \times 10^6$ & 92.5 & $2.60 \times 10^7$ \\
DyLAN             & 80.2 & $1.20 \times 10^6$ & 88.2 & $1.40 \times 10^7$ \\
Vanilla           & 82.1 & $1.50 \times 10^5$ & 85.4 & $3.50 \times 10^6$ \\
\bottomrule
\end{tabular}%
} % <- 正确闭合 \resizebox
\end{table}

\section{Experiment}

\subsection{Experimental Setup}

To evaluate Agent-GSPO, we compare it against a comprehensive suite of single-agent and multi-agent baselines. Single-agent methods include Vanilla direct prompting, Chain-of-Thought (CoT)\cite{CoT}, Complex CoT\cite{complexcot}, and Self-Consistency (SC)\cite{sc}. Multi-agent system (MAS) methods feature PHP\cite{php}, LLM-Debate\cite{du2023improvingfactualityreasoninglanguage}, DyLAN\cite{Dylan}, and the current state-of-the-art in communication pruning, AgentPrune-R\cite{AgentPrune-R}.

Our evaluation spans seven challenging benchmarks: \textbf{MMLU}\cite{MMLU} for general reasoning; \textbf{GSM8K}\cite{gsm8k}, \textbf{MultiArith}\cite{MultiArith}, \textbf{SVAMP}\cite{SVAMP}, \textbf{AQuA}\cite{AQuA}, and \textbf{MATH-500}\cite{MATH} for mathematical reasoning; and \textbf{HumanEval}\cite{humaneval} for code generation. To ensure a fair comparison, all experiments leverage \texttt{gpt-4-1106-preview} as the base model for all agents, maintaining consistency with prior work. The experimental setup requires collaboration for task success, as essential information is distributed among agents, making efficient communication critical. For specific hyperparameters, please refer to the Appendix. All experiments were conducted on NVIDIA A100 80GB GPUs.

\subsection{Performance Analysis}
Agent-GSPO establishes a new state-of-the-art across seven reasoning benchmarks, demonstrating superior performance and efficiency. The framework achieves top-tier accuracy with \textbf{96.02\% on GSM8K} and a \textbf{90.70\% pass@1 rate on HumanEval}, alongside a significant \textbf{+9.38 point} absolute gain on the challenging MATH-500 benchmark, reaching \textbf{83.10\%}. Critically, Agent-GSPO's primary contribution lies in its exceptional communication efficiency. It secures these state-of-the-art results on GSM8K while consuming only \textbf{7.2 million tokens}, a stark contrast to the 22-26 million tokens used by competing methods for lower accuracy. This efficiency is consistent across all benchmarks, such as on MMLU where it requires only \textbf{0.92 million tokens}. Results show that optimizing a communication-aware reward enables Agent-GSPO to learn cost-effective strategies that prune verbosity, yielding more accurate, scalable, and economical collaboration.
\subsection{Ablation Study}

\begin{figure}[t]
    \centering
    \includegraphics[width=\linewidth]{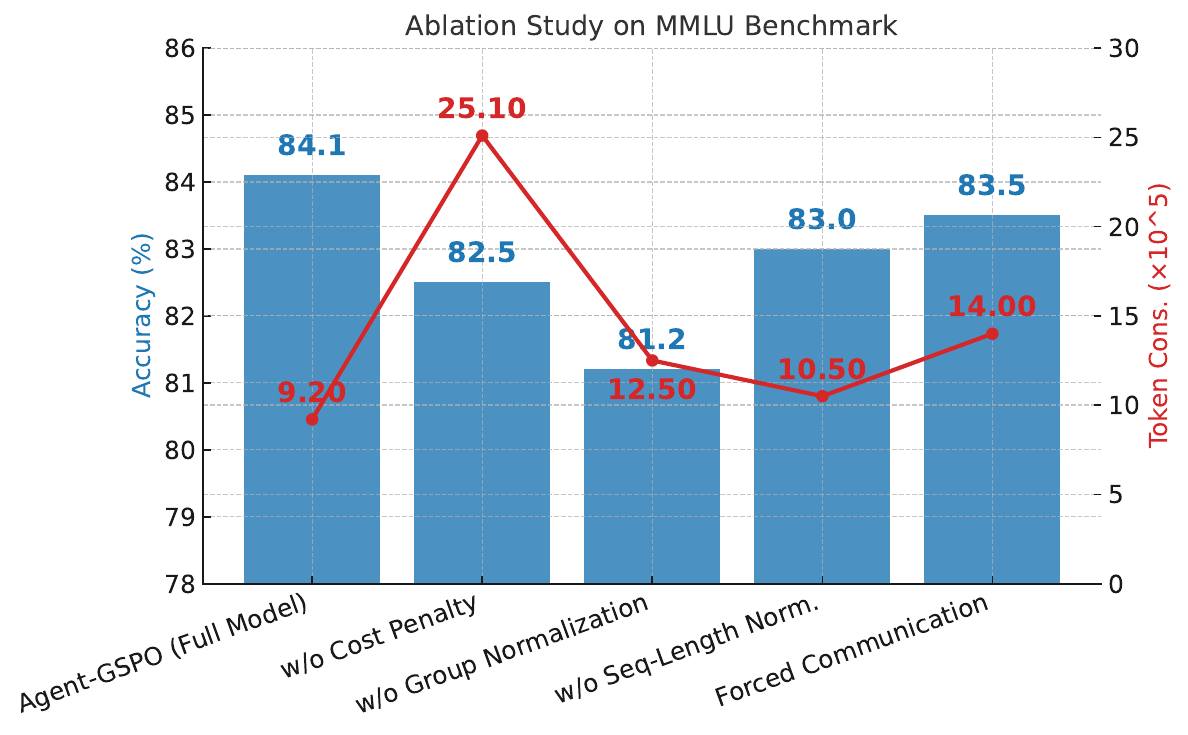}
    \vspace{-0.7cm}
    \caption{Ablation study on the MMLU benchmark.}
    \label{fig:ablation_mmlu}
\end{figure}

To isolate the contributions of our core components, we conducted an ablation study on the MMLU benchmark, with results presented in figure~\ref{fig:ablation_mmlu}. The analysis confirms our central hypothesis: removing the communication cost penalty ($\lambda_{\text{tok}}=0$) causes token consumption to nearly triple while degrading accuracy, validating the need for an explicit economic disincentive. Furthermore, disabling GSPO's group-wise advantage normalization results in the most significant accuracy drop ($\downarrow$2.9 points), underscoring its critical role in stabilizing the high-variance, sequence-level training process.
\section{Conclusion}
\label{sec:conclusion}

In this paper, we introduced \textbf{Agent-GSPO}, a novel framework that addresses communication inefficiency in multi-agent systems by framing it as a sequence-level reinforcement learning problem. By directly optimizing a communication-aware reward using the stable and efficient Group Sequence Policy Optimization (GSPO) algorithm, our method trains agents to balance task performance with token economy. Our experiments demonstrate that Agent-GSPO establishes a new state-of-the-art across seven reasoning benchmarks with a fraction of the communication cost of prior methods. This efficiency stems from the emergent skill of ``strategic silence," where agents prune verbosity, paving the way for scalable, cost-effective, and rational multi-agent systems.
\bibliographystyle{IEEEbib}
\bibliography{strings,refs}

\end{document}